\documentclass[12pt]{article}%
\usepackage[utf8]{inputenc}%
\usepackage[english]{babel}%
\usepackage[dvips]{graphicx}%
\usepackage{hhline}%
\usepackage{longtable}%
\begin{document}

\begin{center}
{\bf \large BLUESHIFTED GALAXIES IN THE VIRGO CLUSTER}\\
\bigskip
I. D. Karachentsev and O. G. Nasonova (Kashibadze)
\bigskip
\end{center}

{\itshape{}We examine a sample of 65 galaxies in the Virgo cluster
with negative radial velocities relative to the Local Group.
Some features of this sample are pointed out. All of these
objects are positioned compactly within a virial zone of
radius 6$^{\circ}$ in the cluster, but their centroid is
displaced relative to the dynamic center of the cluster,
M\,87, by 1$^{\circ}\hspace{-0.4em}.\,$1 to the northwest. The dwarf galaxies in
this sample are clumped on a scale of $\sim$10' (50 kpc). The
observed asymmetry in the distribution of the blueshifted
galaxies may be caused by infall of a group of galaxies
around M\,86 onto the main body of the cluster. We offer
another attempt to explain this phenomenon, assuming a
mutual tangential velocity of $\sim$300 km s$^{-1}$ between the
Local Group and the Virgo cluster owing to their being
repelled from the local cosmological void.}

{\bf{}Keywords}: galaxies: blueshifted: Virgo

\section{Introduction}

Of the approximately one million galaxies with measured radial
velocities, about a hundred of them have negative radial velocities
relative to the center of the Local Group. If we exclude the 31
galaxies which form part of the Local Group (LG), the rest are
distributed over the sky in an extremely nonuniform fashion:
two dwarf spheroidal galaxies (KK\,77 and IKN) with velocities
of --96 and  --1 ($\pm60)$ km s$^{-1}$, respectively,
are satellites of the neighboring M\,81 spiral, while the
remaining 65 galaxies are concentrated in the central region
of the nearest cluster in Virgo around the giant elliptical
galaxy Virgo A = NGC\,4486. It is possible that several other
blueshifted galaxies exist in the group nearest us around the
giant galaxies Maffei 1, Maffei 2, and IC\,342, but it is very
difficult to measure radial velocities in this region of the
sky because of the strong ($\sim5^m - 6^m)$ absorption and emission
by hydrogen in the Galaxy. It is evident that the observed
distribution over the sky of galaxies with negative radial
velocities reflects
both the depth and the proximity to us of neighboring potential
wells; that is, it contains important information on the local
segment of the evolving large-scale structure.

\section{List of blueshifted galaxies in Virgo}

According to the Virgo Cluster Catalog [1] the population of the
cluster numbers more than 2000 members, most of which are dwarf
irregular (dIr), elliptical (dE), and spheroidal (dSph) systems.
Their overall number has increased with time as a result of
different surveys of galaxy redshifts in the northern sky, as
well as because of special programs aimed at studying the
kinematics of the Virgo cluster [2--9]. As Binggeli et al. [7],
as well as others, have shown the Virgo cluster consists of
several subclusters which differ in their average velocities,
dispersions in their velocities, and dominant types in their
populations; this suggests that the dynamic relaxation of the
cluster is incomplete. The bulk of the population of Virgo is
concentrated around the brightest galaxy NGC\,4486 (M\,87). The
X-ray intensity peak of the hot intergalactic gas in Virgo
also lies in M\,~87, which suggests that this radio galaxy is
the dynamic center of the cluster. According to Binggeli
et al. [7], the average heliocentric velocity of the main
Virgo cluster is +1050 $\pm$ 35 km s$^{-1}$ with a standard
deviation of $\sigma = 650$ km s$^{-1}$. Here the average and
standard deviation depend significantly on the choice
of boundaries for the cluster and the morphological
type of the galaxies.

\begin{figure}[b]
\begin{center}
\includegraphics[width=0.8\textwidth,keepaspectratio]{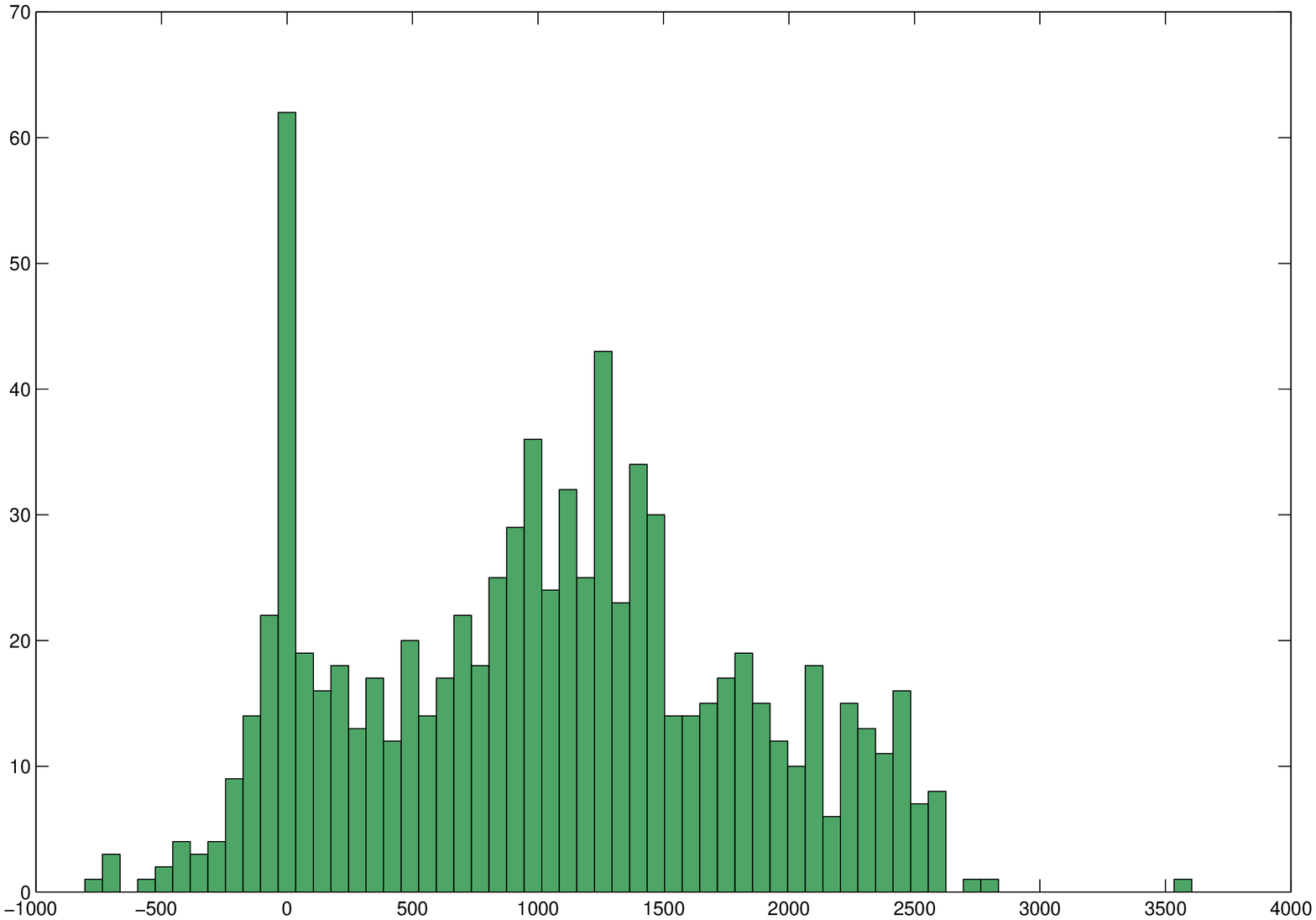}
\vbox{\vspace{0.6cm}
\parbox{0.8\textwidth}{\footnotesize{}%
Fig. 1. The distribution of the measured radial velocities of the
825 galaxies in the vicinity of the Virgo cluster. The peak near
zero velocity is caused by ``star spam'' from the
SDSS survey.}}
\end{center}
\end{figure}

At the time [7] was published,
the total number of galaxies with negative radial velocities
relative to the centroid of the Local Group ($V_{LG} < 0)$ was 42.
For some of the galaxies (VCC\,584, VCC\,1035) the negative
velocities have not been confirmed by subsequent, more
accurate measurements. A great advance in the study of
the kinematics was provided by a ``blind''
survey of the cluster in the HI line at the Arecibo radio
telescope [10--12], as well as by the appearance of new
radial velocity data from the Sloan Digital Sky Survey
SDSS [13]. Here it should be noted that the SDSS survey
yielded improved values for the radial velocities of the
members of Virgo, but it also produced
a number of false ``galaxies'' with velocities $V_h\simeq 0$.
Fig.~1 shows the distribution of the radial velocities of
$\sim$800  galaxies in the central region of Virgo. A sharp peak
can be seen near $V_h\simeq 0$ owing mainly to binary stars
which are, nevertheless, listed in the HyperLeda data base as galaxies.
It is evident that more than one hundred of these observations can
 significantly distort the statistics of blueshifted galaxies in
Virgo.

{\footnotesize
\begin{center}
Table 1. List of Virgo cluster galaxies with negative radial velocities.
\end{center}
\vspace{-0.5cm}
\begin{longtable}{|l|l|rr|r|l|l|l|}
\hhline{|=|=|==|=|=|=|=|}
Designation& ~~~~RA (J2000.0) Dec &  $V_h$ &$\pm$&$V_{LG}$& Type& ~$B_T$&Notes              \\
\multicolumn{1}{|c|}%
{(1)}&\hspace{5eM}(2)&\multicolumn{2}{c|}{~~~~(3)}&   (4)  & (5) &  ~(6) &(7)                \\
\hhline{|=|=|==|=|=|=|=|}\endfirsthead
\hhline{|=|=|==|=|=|=|=|}
\multicolumn{1}{|c|}%
{(1)}&\hspace{5eM}(2)&\multicolumn{2}{c|}{~~~~(3)}&   (4)  & (5) &  ~(6) &(7)                \\
\hhline{|=|=|==|=|=|=|=|}\endhead
\hhline{|=|=|==|=|=|=|=|}\endfoot
\hhline{|=|=|==|=|=|=|=|}\endlastfoot
IC3036  & 12 12 15.08 +12 29 17.7 &   $-$2 &  38 & $-$126 & Sm  & 14.66 &                   \\
IC3044  & 12 12 48.49 +13 58 34.6 & $-$182 &   4 & $-$298 & SBc & 14.23 &                   \\
VCC087  & 12 13 41.27 +15 27 13.0 & $-$159 &   2 & $-$267 & Sm  & 15.20 &                   \\
NGC4192 & 12 13 48.29 +14 54 01.6 & $-$135 &   7 & $-$246 & Sb  & 10.85 &                   \\
NGC4212 & 12 15 39.38 +13 54 05.7 &  $-$84 &   4 & $-$199 & Sc  & 11.78 &                   \\
VCC181  & 12 16 14.63 +13 35 11.6 & $-$150 &  32 & $-$267 & Im  & 17.42 &                   \\
VCC200  & 12 16 33.71 +13 01 53.1 &     22 &  24 &  $-$98 & E   & 14.80 &                   \\
A224385 & 12 16 49.76 +13 30 14.4 &  $-$87 &   4 & $-$204 & BCD & 18.42 &                   \\
IC3094  & 12 16 56.03 +13 37 31.6 & $-$159 &   6 & $-$275 & Sab & 14.43 &                   \\
VCC237  & 12 17 29.40 +14 53 10.0 & $-$323 &   5 & $-$423 & Sdm & 16.79 &                   \\
IC3105  & 12 17 33.72 +12 23 17.6 & $-$162 &   3 & $-$284 & Sm  & 14.75 &                   \\
VCC322  & 12 19 05.05 +13 58 52.1 & $-$209 &  10 & $-$323 & IAB & 15.36 & Foreground?       \\
VCC334  & 12 19 14.27 +13 52 56.1 & $-$236 &   4 & $-$350 & BCD & 16.14 &                   \\
VCC501  & 12 21 47.99 +12 49 36.1 & $-$105 &  33 & $-$224 & E   & 17.29 & Noisy SDSS Sp     \\
IC3224  & 12 22 36.12 +12 09 27.0 &     10 &   5 & $-$100 & BCD & 16.6  &                   \\
VCC628  & 12 23 15.47 +07 41 22.2 & $-$398 &  10 & $-$540 & Ir  & 18.43 &                   \\
VCC636  & 12 23 21.23 +15 52 06.2 &   $-$9 &  28 & $-$113 & S0  & 16.48 &                   \\
IC3258  & 12 23 44.45 +12 28 39.4 & $-$473 &   6 & $-$593 & IB  & 13.66 & Merger?           \\
IC3303  & 12 25 15.20 +12 42 52.3 & $-$309 &  25 & $-$427 & E   & 14.75 &                   \\
VCC788  & 12 25 16.82 +11 36 19.5 &    121 &  29 &   $-$3 & E   & 16.79 &                   \\
VCC802  & 12 25 29.06 +13 29 53.6 & $-$204 &   3 & $-$318 & Ir  & 17.61 &                   \\
IC3311  & 12 25 33.06 +12 15 37.8 & $-$166 &  10 & $-$287 & Scd & 14.70 &                   \\
VCC810  & 12 25 33.56 +13 13 38.1 & $-$354 &  30 & $-$470 & E   & 16.63 & Noisy SDSS Sp     \\
VCC815  & 12 25 37.20 +13 08 37.7 & $-$750 &  27 & $-$866 & E   & 16.33 &                   \\
VCC846  & 12 25 50.56 +13 11 51.8 & $-$729 &  30 & $-$845 & E   & 16.20 &                   \\
NGC4396 & 12 25 58.94 +15 40 16.6 & $-$112 &   5 & $-$215 & Scd & 13.09 &                   \\
VCC877  & 12 26 09.56 +13 40 23.3 &  $-$99 &  56 & $-$212 & E   & 17.68 &                   \\
NGC4406 & 12 26 11.69 +12 56 46.0 & $-$258 &  23 & $-$374 & E   &  9.83 &                   \\
VCC892  & 12 26 20.04 +12 30 36.6 & $-$666 &  68 & $-$784 & E:  & 18.45 &                   \\
NGC4413 & 12 26 32.21 +12 36 38.6 &    103 &   5 &  $-$16 & Sab & 12.87 &                   \\
VCC928  & 12 26 39.80 +12 30 48.2 & $-$276 &  23 & $-$395 & E   & 16.23 &                   \\
IC3355  & 12 26 51.29 +13 10 27.8 &  $-$10 &   6 & $-$126 & IB  & 15.41 & Foreground?       \\
VCC953  & 12 26 54.74 +13 33 58.3 & $-$450 &  30 & $-$563 & E   & 15.91 &                   \\
NGC4419 & 12 26 56.44 +15 02 50.9 & $-$277 &   8 & $-$383 & SBa & 11.99 &                   \\
VCC997  & 12 27 22.18 +12 04 07.5 & $-$240 & 118 & $-$360 & E   & 18.25 & Noisy SDSS Sp     \\
KDG132  & 12 27 31.64 +09 36 08.6 &     32 &  33 & $-$100:& Ir  & 16.43 & SDSS Sp for a knot\\
NGC4438 & 12 27 45.58 +13 00 31.8 &     73 &   8 &  $-$43 & Sa  & 10.93 &                   \\
SDSS    & 12 28 25.86 +11 14 25.1 &    124 &  50 &   $-$0 & E   & 18.25 &                   \\
VCC1129 & 12 28 44.98 +12 48 35.7 &     12 & 138 & $-$105 & E   & 17.75 &                   \\
VCC1163 & 12 29 06.43 +14 00 18.5 & $-$453 &  26 & $-$564 & E   & 16.56 &                   \\
VCC1175 & 12 29 18.20 +10 08 09.2 &     11 &   3 & $-$118 & BCD & 15.37 &                   \\
VCC1198 & 12 29 32.06 +13 30 37.8 & $-$357 &  37 & $-$470 & E   & 17.82 &                   \\
IC3416  & 12 29 34.98 +10 47 35.8 &  $-$72 &  41 & $-$198 & Ir  & 15.04 &                   \\
VCC1239 & 12 29 51.18 +13 52 04.6 & $-$561 &  28 & $-$672 & E   & 15.68 &                   \\
VCC1264 & 12 30 10.91 +12 11 44.1 & $-$420 &  59 & $-$539 & E   & 16.90 &                   \\
IC3435  & 12 30 39.85 +15 07 47.3 &  $-$45 &  22 & $-$150 & S0  & 15.53 &                   \\
VCC1314 & 12 30 49.03 +13 13 26.1 &     77 &  40 &  $-$37 & E   & 17.34 &                   \\
IC3445  & 12 31 19.42 +12 44 16.9 & $-$354 &  23 & $-$470 & E   & 16.49 &                   \\
IC3471  & 12 32 22.85 +16 01 08.3 & $-$135 &   2 & $-$235 & Sdm & 15.47 &                   \\
IC3476  & 12 32 41.82 +14 03 04.0 & $-$170 &   7 & $-$280 & Sdm & 13.36 &                   \\
IC3492  & 12 33 19.80 +12 51 12.8 & $-$489 &  25 & $-$604 & E   & 14.73 &                   \\
IC3548  & 12 35 56.62 +10 56 10.9 &     86 &  28 &  $-$37 & E   & 16.08 &                   \\
VCC1682 & 12 36 36.72 +14 13 32.8 &     41 &  36 &  $-$66 & E   & 17.86 &                   \\
NGC4569 & 12 36 49.86 +13 09 48.1 & $-$233 &   4 & $-$345 & Sa  & 10.11 &                   \\
UGC7795 & 12 37 45.34 +07 06 14.0 &     62 &   3 &  $-$78 & Ir  & 14.72 & Foreground?       \\
VCC1750 & 12 38 15.54 +06 59 39.1 & $-$117 &  10 & $-$258 & BCD & 16.76 &                   \\
VCC1761 & 12 38 27.74 +14 04 38.2 & $-$162 &  27 & $-$269 & E   & 16.95 &                   \\
KDG172  & 12 39 13.86 +15 37 49.4 &     57 &  10 &  $-$42 & Ir  & 17.61 &                   \\
VCC1812 & 12 39 55.55 +11 51 28.5 & $-$234 &  41 & $-$351 & E   & 17.31 &                   \\
VCC1860 & 12 40 57.29 +15 16 31.1 &  $-$24 &  40 & $-$124 & E   & 18.12 & Noisy SDSS Sp     \\
IC3658  & 12 41 20.65 +14 42 02.4 &     34 &  20 &  $-$69 & E   & 14.94 &                   \\
UGC7857 & 12 41 54.24 +13 46 22.8 &    101 &  31 &   $-$7 & E   & 14.72 &                   \\
VCC1909 & 12 42 07.45 +11 49 42.0 &    101 &  38 &  $-$16 & E   & 16.16 &                   \\
IC0810  & 12 42 09.11 +12 35 48.8 &  $-$75 &  23 & $-$188 & S0  & 14.25 &                   \\
VCC2028 & 12 45 37.48 +13 19 42.8 &     56 &  28 &  $-$52 & E   & 16.72 &                   \\
\end{longtable}
}

We have made a detailed analysis of the available
observational data and, by excluding the ``star spam'',
compiled a list of 65 galaxies in Virgo with $V_{LG} < 0$. These are
listed in Table 1. The columns of the table list the following
data on the galaxies: (1)~galaxy number in standard catalogs,
(2)~equatorial coordinates at the epoch J2000.0, (3)~average
heliocentric radial velocity and the error in it, (4)~radial
velocity relative to the centroid of the Local Group with the
apex parameters from [14] used in the NED, (5)~morphological
type, (6)~B-band apparent total magnitude, and (7)~brief comments.
Of the 65 galaxies in this list, the radial velocities of 27 have
been determined with high accuracy using the HI line.

\begin{figure}
\begin{center}
\includegraphics[width=0.8\textwidth,keepaspectratio]{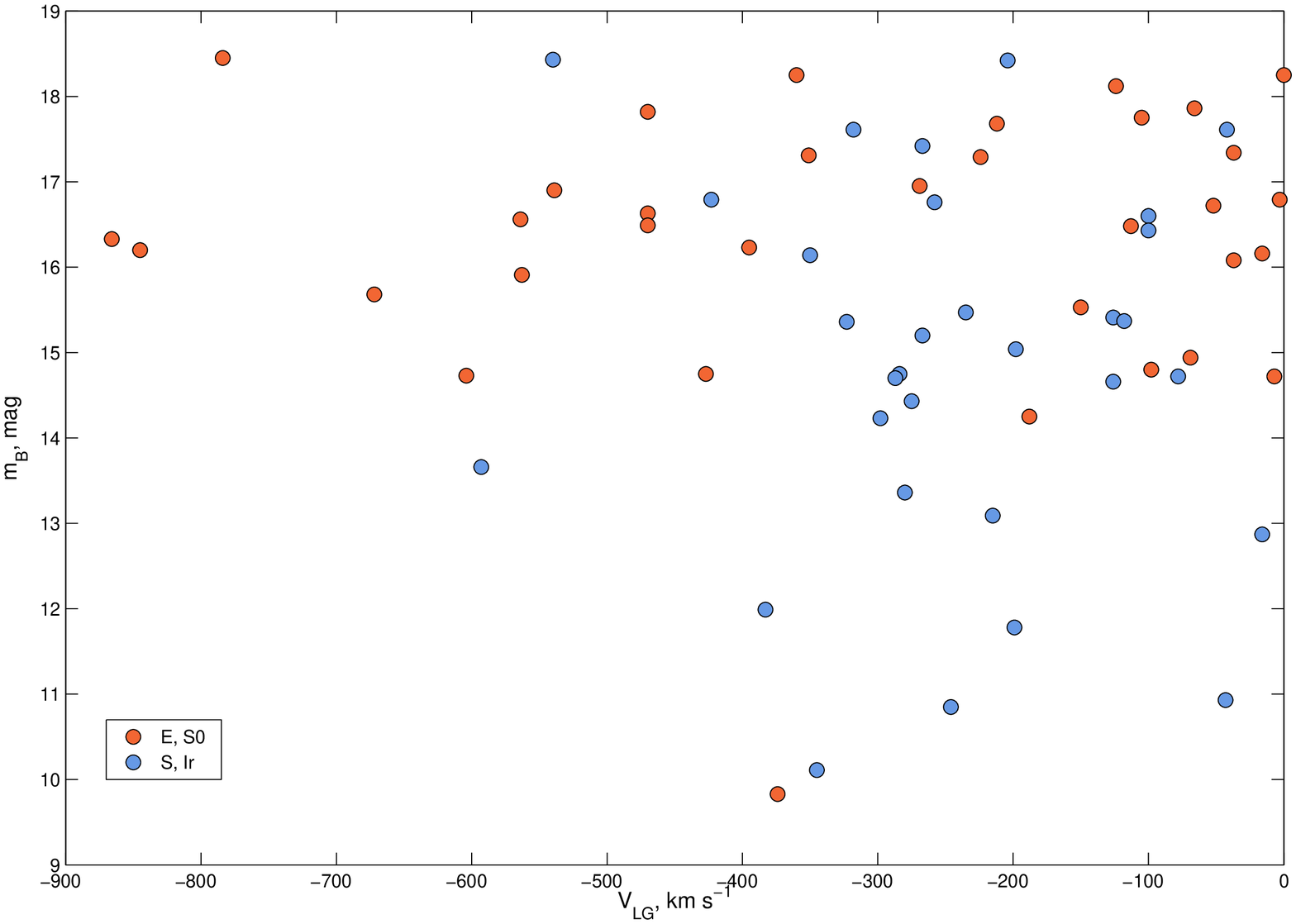}
\vbox{\vspace{0.6cm}
\parbox{0.8\textwidth}{\footnotesize{}%
Fig. 2. Apparent magnitudes and radial velocities of galaxies
in Virgo that are approaching us. Galaxies of early (E, S0,
dSph) and late (S, Ir, BCD) types are indicated, respectively,
by red and blue circles.}}
\end{center}
\end{figure}

\begin{figure}
\begin{center}
\includegraphics[width=0.8\textwidth,keepaspectratio]{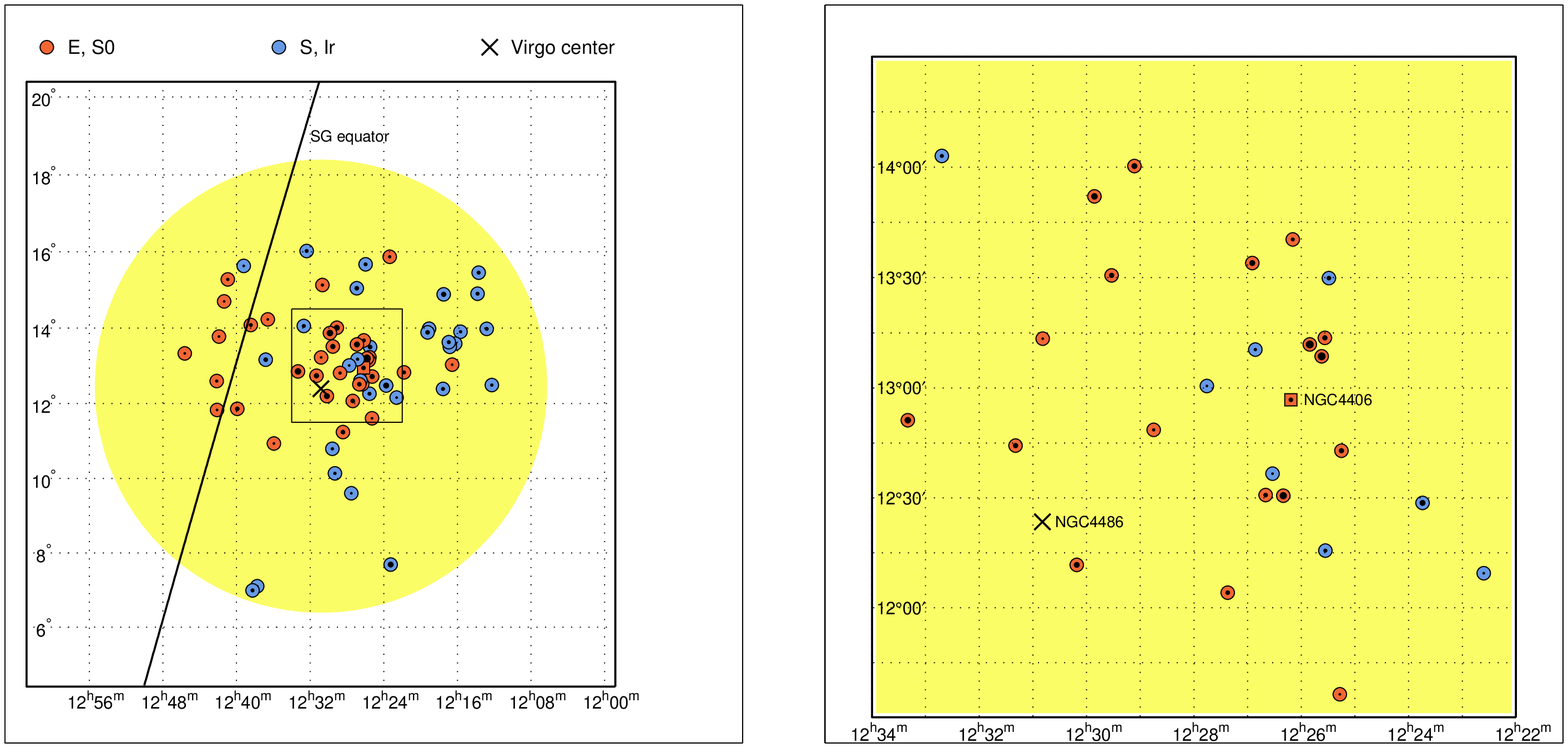}
\vbox{\vspace{0.6cm}
\parbox{0.8\textwidth}{\footnotesize{}%
Fig. 3. The distribution in equatorial coordinates of the 65
galaxies with negative radial velocities. The circle, of radius
6$^{\circ}$, filled with yellow color
marks the virial zone of the cluster around the radio
galaxy M\,87, which is indicated by an X. The densest region
(the square) is magnified in the frame on the right.}}
\end{center}
\end{figure}

The distribution of the blueshifted galaxies with respect to
their radial velocities $V_{LG}$ and apparent magnitudes
$m_B$ is shown in Fig.~2. The galaxies of early types (E, S0, dE,
dSph) are indicated by red circles, and those with a young population (S, dIr,
BCD) are indicated by blue ones. For a distance to the Virgo cluster
of 17.0 Mpc [15] and an absolute magnitude for the dwarf galaxies
fainter than --16.5$^m$, their relative number in the sample is 80\,\%.
Of the 13 galaxies with normal and high luminosities, only one,
NGC\,4406, is elliptical.

Fig.~3 shows the distribution of the
galaxies with negative velocities over the sky in equatorial
coordinates. The virial region of the cluster, with a radius
of $\Theta_{VIR} = 6^{\circ}\hspace{-0.4em}.\,0$, is indicated by a large circle. The galaxies
of earlier and later types are indicated in the same way as in the
previous figure. The position of the galaxy M\,87 as the physical
center of the cluster is indicated by a cross. The inclined
straight line corresponds to the supergalactic equator. The
thick central zone, itself, is shown enlarged in the right hand
frame of the figure.

\section{Some features of the distribution of blueshifted galaxies}

The distribution over the sky of the 65 galaxies with negative
radial velocities is characterized by several features that may
be important for understanding the kinematics and evolution of
the cluster.

 (a)~The blueshifted galaxies are distributed over
the sky much more compactly than the rest of the population of
Virgo. All the galaxies with $V_{LG} < 0$ lie, without exception,
inside the virial radius $\Theta_{VIR}=6^{\circ}\hspace{-0.4em}.\,0$ of the cluster.

 (b)~The centroid of the present sample does not coincide with the
dynamic center of Virgo (M\,87), and is shifted significantly to
the NW by $1^{\circ}\hspace{-0.4em}.\,10\pm 0^{\circ}\hspace{-0.4em}.\,35$.

 (c)~The galaxies of early and late
types have significantly different positions relative to the
center of the cluster: E and SO galaxies lie predominantly to
the east, while objects with a young population (dIr, BCD, S)
lie mainly to the west and south.

 (d)~The dwarf galaxies with
$V_{LG} < 0$ manifest a small scale clumping effect. Thus, the
galaxies VCC\,181, AGC\,224385, and IC\,3094 form a triplet in projected
onto the sky, while VCC\,322 and VCC\,334 form a pair with a small
difference in radial velocities. Other examples of multiple systems
are VCC\,810/815/846, VCC\,892/928, and UGC\,7795/VCC\,1750. The
characteristic scale of the clumping for these objects is $\sim10$'
or 50 kpc, while the median difference in their velocities is
about 70 km s$^{-1}$. These groups contain galaxies with young, as
well as old, populations. If it is real, the dynamic isolation
of these pairs and triplets within the cluster is highly unexpected.

\section{Discussion}

Binggeli et al. [7] and Jerjen et al. [16] proposed an explanation
for the observed shift of the centroid of the galaxies with negative
velocities relative to the dynamic center of Virgo in terms of a
separate grouping of galaxies associated with the giant elliptical
galaxy NGC\,4406 (M\,86). In the sky this galaxy lies near the
centroid of the blueshifted galaies (see the right hand frame
of Fig.~3), and its radial velocity, $-$374 km s$^{-1}$, is close to
the average velocity of our sample of galaxies (see Fig.~2).
According to these authors, M\,86, together with the cloud of
satellites surrounding it, is incident onto (merging with) the
center of Virgo (M\,87) from the further edge of the cluster at
a relative velocity of $\sim 1400$ km s$^{-1}$, which is a bit more than twice
the mean square virial velocity of the members of Virgo. Moving at
this velocity, M\,86 and its companions will intersect the virial
radius of the cluster (1.8 Mpc) after 1.3 billion years and will
continue moving towards our Galaxy.

The distances to M\,86 and six
other galaxies with negative radial velocities, VCC\,200, VCC\,810,
VCC\,815, VCC\,846, VCC\,928, and NGC\,4419 have been measured [15--17]
from the surface brightness fluctuations (SBF). The distance of
M\,86, itself, 17.5$\pm$0.4 Mpc, and the average distance of the
other six galaxies, 17.3$\pm$1.1 Mpc are the same, to within the
limits of error, as the average for the cluster as a whole (17.0
Mpc).

The hypothesis according to which the grouping of galaxies
around M\,86 is incident on the main body of the cluster centered
at M\,87 seems quite plausible, as it is consistent with the general
paradigm of cluster formation through the merger of small clumps
(groups). However, the reason for the observed spatial segregation
in terms of morphological types among the galaxies of Virgo with
negative radial velocities is still unclear. The scenario in which
the cloud around M\,86 transits through the virial zone of Virgo
does not provide an explanation for the observed bunching of
dwarf galaxies in the cloud of M\,86 on a scale of $\sim$50 kpc, since
low mass multiple systems of this sort could easily be destroyed
by tidal forces in the central zone of the cluster. Numerical
simulation of the passage of an association of dwarf groups through
the central region of the Virgo cluster might help resolve this
puzzle.

In addition to the above features, there is yet another
distinct tendency that can be seen in the subsystem of blueshifted
galaxies in Virgo. If we restrict the sample to ever more negative
values of $V_{LG}$, then the position in the sky of the centroid of the
remaining galaxies shifts systematically to the NW, while the
spread in the positions of the galaxies relative to the sliding
centroid becomes smaller. The upper frame of Fig.~4 illustrates
the drift in the sky of the centroid of the galaxies, which range
in terms of $V_{LG}$ from zero to the maximum negative value (--866
km s$^{-1}$). The positions of the sliding centroid are indicated by
dots which are joined by straight lines. The numbers next to them
correspond to the number of galaxies in the averages and the numbers
in parentheses, to the average radial velocities obtained by
averaging the radial velocities of the 15, 20, 25, ..., 65 members
of the sample. The lower frame of Fig.~4 shows the analogous drift
of the centroid, but with the reverse ordering, from --866 km s$^{-1}$
to zero. The behavior of the sliding centroid in the plots of Fig.~4
is obviously difficult to match with a simple
picture where all the blueshifted galaxies form a group around
NGC\,4406 which is incident, as a whole, on the Virgo cluster with
its center alongside NGC\,4486.

\begin{figure}[ht]
\begin{center}
\includegraphics[height=0.8\textwidth,keepaspectratio,angle=270]{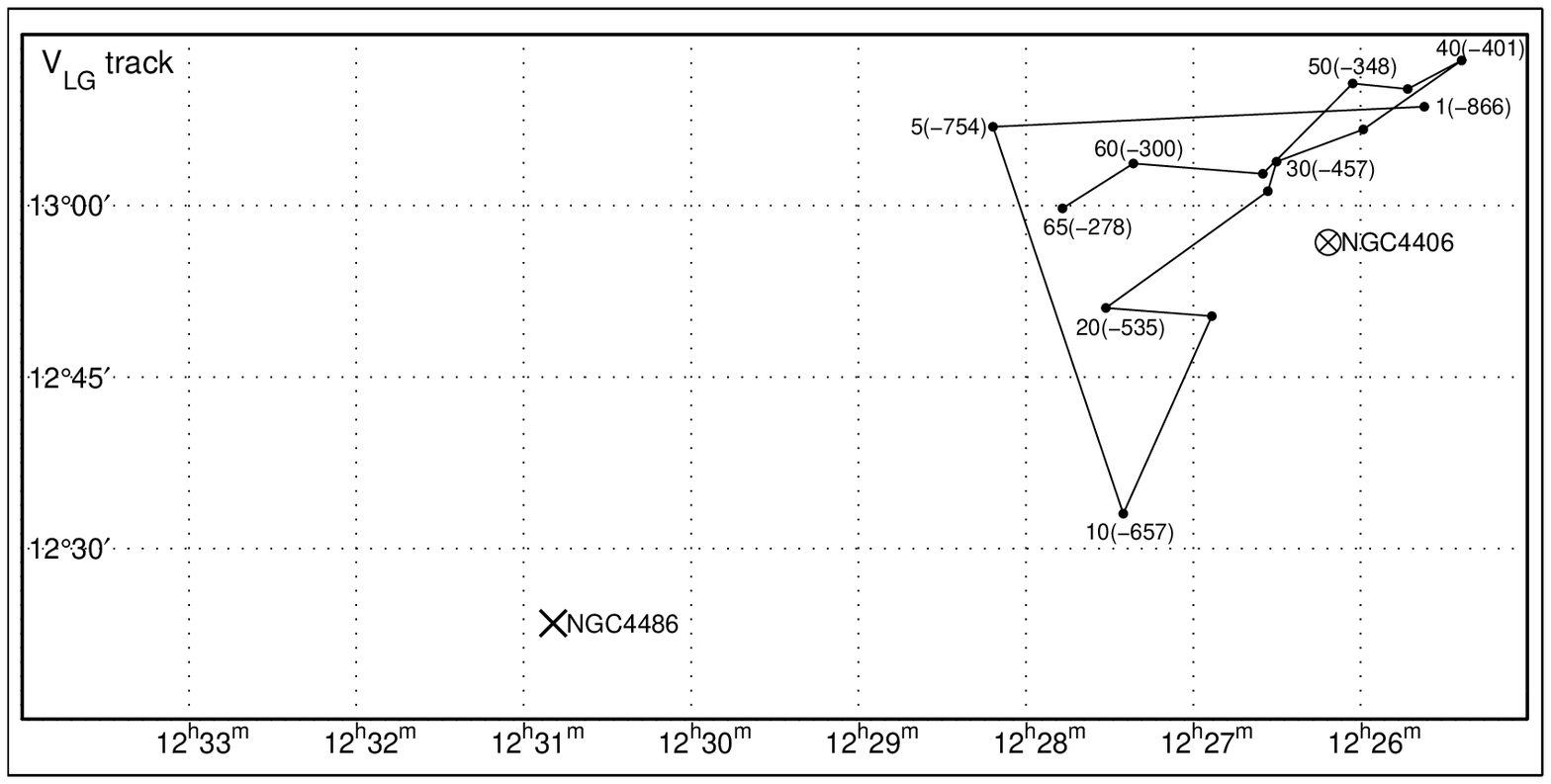}
\includegraphics[height=0.8\textwidth,keepaspectratio,angle=270]{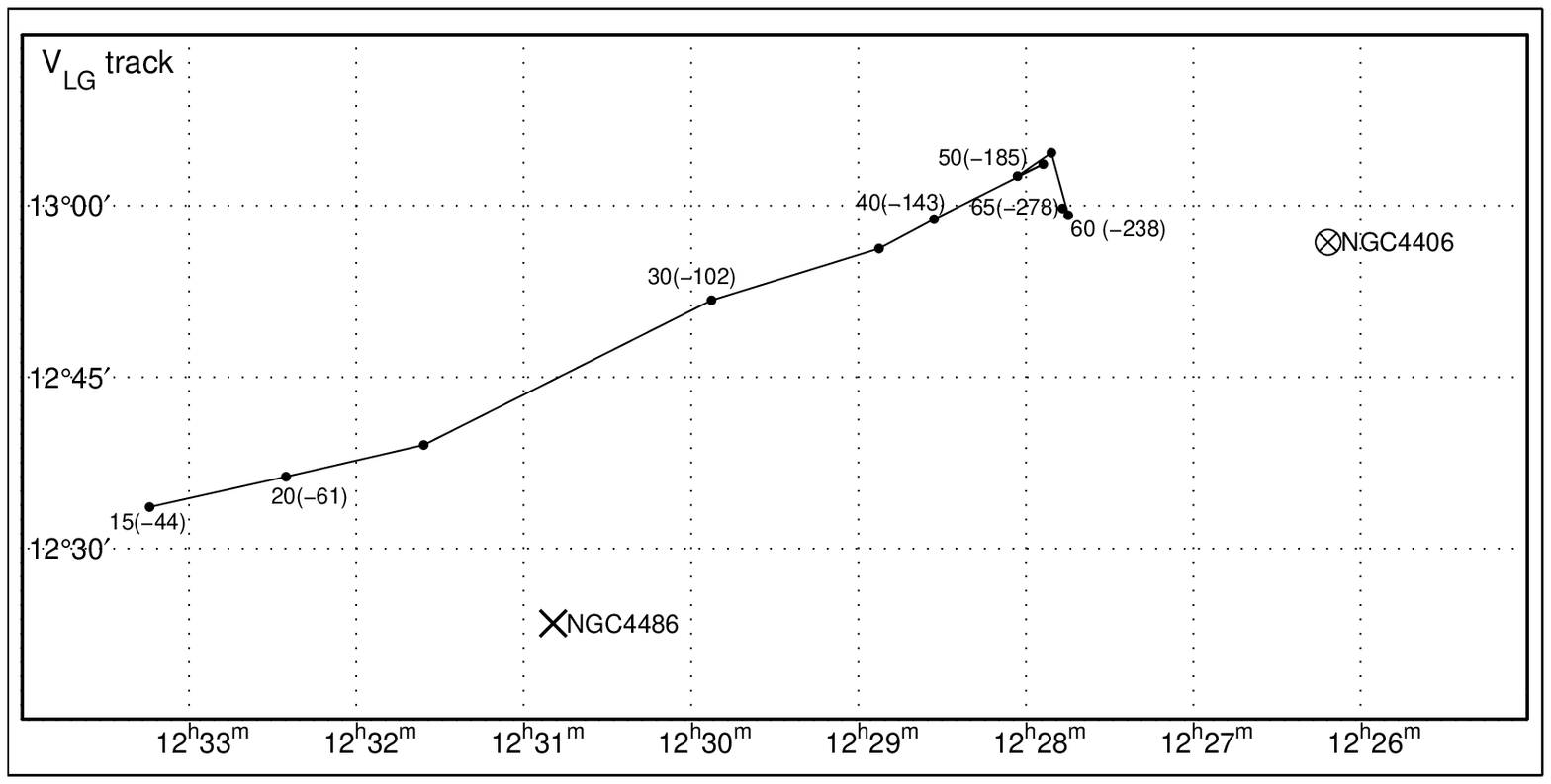}
\vbox{\vspace{0.6cm}
\parbox{0.8\textwidth}{\footnotesize{}%
Fig. 4. The position of the centroid of the galaxies, arranged
according to the magnitude of their negative radial velocity, as
a function of the number of averaged objects with a step size of
5. The numbers in parentheses are the average velocity in km s$^{-1}$.
The upper and lower frames correspond to two different ways of
averaging.}}
\end{center}
\end{figure}

Another argument can be advanced for
explaining the observed features of the galaxies in Virgo with
$V_{LG} < 0$. Let the kinematics of the cluster correspond to strictly
radial motion in a spherically symmetric cluster with no
significant substructures. Evidently, the galaxies with the
highest peculiar velocities, both toward and away from the
observer, will be concentrated within the narrowest zone of
the cluster near its physical center. In this picture, the extreme
negative velocities in Virgo should be expected right around M\,87,
where the velocities of the galaxies are directed almost strictly
along the line of sight. But this would be true only if the centroid
of the Local Group does not have a tangential velocity relative to
the Virgo cluster. If the centers of the Local Group and Virgo move
at a mutually tangential velocity $V_{tang}$, then for two galaxies
located at the edges of the virial diameter of Virgo along the
direction of the vector $V_{tang}$, the projection of this component
would lead to a difference of
$\Delta V = 2\times\sin\Theta_{VIR}\times V_{tang}$
in the radial velocities. This in turn, would produce a visible
shift of the centroid of the galaxies selected for their negative
velocities.

In a study of the local peculiar velocity field of galaxies with
measured distances, Tully et al. [18] have determined three main
components of the velocity of the Local Group, which together yield
its observed motion relative to the cosmic microwave background at
a velocity of 630$\pm$20 km s$^{-1}$. One component corresponds to infall
of the Local Group on Virgo at a velocity of 185$\pm$20 km s$^{-1}$, the
second, to recession from the Local Void at a velocity
of 259$\pm$25 km s$^{-1}$, and the third, at 455$\pm$15 km s$^{-1}$ is oriented
toward the massive Great Attractor (GA) in the HydraCentaurus
constellations. These three components are roughly mutually
perpendicular. Since the center of the Virgo cluster lies
almost on the equator of the Local Supercluster, and the Local
Void lies near the supergalactic pole, with $V_{tang}$ = 259 km s$^{-1}$ and
$\theta_{VIR} = 6^{\circ}$ we can expect a gradient of the local velocity
$\Delta V\simeq 54$ km s$^{-1}$ oriented in Fig.~2  at a
right angle rightward of the equator of the local supercluster.
Since the angular size of the infall zone around
 Virgo is $\Theta_0= 23^{\circ}$ [19], the drop in the radial velocities
within it cannot exceed $\Delta V\simeq -183$ km s$^{-1}$. The
magnitude
of the shift in the centroid of the galaxies with $V_{LG} < 0$ induced
by this effect depends on the structural and kinematic features
of Virgo. Dividing the sample of 65 galaxies with $V_{LG} < 0$ into
two halves by a line on the sky parallel to the supergalactic
equator, we obtain an average difference of the radial velocities
between the right and left halves
of $-68\pm51$ km s$^{-1}$. When the sample is split into two by a
line perpendicular to the line joining the galaxies M\,86 and M\,87,
this difference increases to $-139\pm49$ km s$^{-1}$. Thus, we have rough
agreement between the expected and observed effect of a tangential
velocity in terms of both direction and magnitude. Including the
other component of the LG velocity, directed toward the Great
Attractor, can improve this agreement once we note that the Virgo
cluster (closer to the GA) should be incident on the GA at a higher
velocity than the Local Group. This mutual motion produces an
additional component of the tangential velocity of the LG relative
to Virgo, directed toward the opposite side from the GA (i.e., to
the NNW in Fig.~2). Of course, the roles of the various components
of the mutual motion of the LG and Virgo require more detailed
study.

\section{Conclusions}

Research on the distribution of the members of the Virgo cluster
with extremely high peculiar velocities directed toward or away
from us is an important, but still little used tool
for understanding the structure and kinematics of this closest
cluster. In principle, for choosing between a picture of the merger
of individual groups of galaxies with the main body of the cluster
or an effect involving mutually tangential motion of the Local
Group and Virgo, one could also draw on galaxies with high positive
velocities ($\sim2000-2500$ km s$^{-1}$). However, the distant background of
Virgo includes many galaxies with such velocities, so it would be
difficult to interpret the observational data.

It should be
emphasized that the faintest galaxies in our sample have absolute
magnitudes MB brighter than $-12.5^m$. We might expect that the cluster
contains a still larger number of faint galaxies, including some
with negative radial velocities. An increase in their statistics
is extremely desirable for more detailed analysis of the kinematics
of Virgo.

It is also important to note the need to measure the
individual distances for objects with $V_{LG} < 0$ in Virgo. Here the
most promising method for estimating the distances continues to
be the use of surface brightness fluctuations (SBF) of dE and dSph
galaxies. The Tully-Fisher method which is applied to dIr and BCD
galaxies does not, unfortunately, provide the accuracy needed to
distinguish objects at the front and back edges of the cluster.
For them, the only bulk method is to determine the distances from
the luminosity of the tip of the red giant branch with the aid of
the HST and other orbital telescopes.

There is some interest in
searching for ultracompact dwarf galaxies with negative radial
velocities in Virgo. Judging from the scenarios discussed in the
literature [20, 21], dwarf galaxies of this type are formed as a
result of their spending a long time in the densest virial zone
of a cluster. Thus, observation of even one ultracompact dwarf
with $V_{LG} < 0$ would counter the idea that a loose cloud of galaxies
surrounding M\,86 is incident on (merging with) the main body of
the cluster.

Further observational effort must obviously be
supplemented by numerical simulations of the kinematics of those
members of Virgo with extremely high peculiar velocities.

This
work was supported by grants RFBR 07--02--00005 and RFBR-DFG
06--02--04017.

\section*{REFERENCES}

{\sloppy

\pagebreak[3]\hangindent=1cm\noindent1. B. Binggeli, A. Sandage, and G. A. Tammann, {\itshape{}Astron. J.} {\bf90}, 1681 (1985) (VCC).

\pagebreak[3]\hangindent=1cm\noindent2. I. D. Karachentsev and V. E. Karachentseva, {\itshape{}Letters to Astron. Zh.} {\bf8}, 198 (1982).

\pagebreak[3]\hangindent=1cm\noindent3. G. D. Bothun and J. R. Mould, {\itshape{}Astrophys. J.} {\bf324}, 123 (1988).

\pagebreak[3]\hangindent=1cm\noindent4. G. L. Hoffman, J. Glosson, G. Helou, E. E. Salpeter, and A. Sandage, {\itshape{}Astrophys. J. Suppl. Ser.} {\bf63}, 247 (1987).

\pagebreak[3]\hangindent=1cm\noindent5. G. L. Hoffman, H. L. Williams, E. E. Salpeter, A. Sandage, and B. Binggeli, {\itshape{}Astrophys. J. Suppl. Ser.} {\bf71}, 701 (1989).

\pagebreak[3]\hangindent=1cm\noindent6. M. J. Drinkwater, M. J. Currie, C. K. Young, et al., {\itshape{}Mon. Notic. Roy. Astron. Soc.} {\bf279}, 595 (1996).

\pagebreak[3]\hangindent=1cm\noindent7. B. Binggeli, C. Popescu, and G. A. Tammann, {\itshape{}Astron. Astrophys.}, {\bf98}, 275 (1993).

\pagebreak[3]\hangindent=1cm\noindent8. G. Gavazzi, C. Bonfanti, P. Pedotti, et al., {\itshape{}Astron. Astrophys.} {\bf146}, 259 (2000).

\pagebreak[3]\hangindent=1cm\noindent9. J. M. Solanes, T. Sanchis, E. Salvador-Sole, R. Giovanelli, and M. P. Haynes, {\itshape{}Astron. J.} {\bf124}, 2440 (2002).

\pagebreak[3]\hangindent=1cm\noindent10. R. Giovanelli, M. P. Haynes, B. R. Kent, et al., {\itshape{}Astron. J.} {\bf133}, 2569 (2007).

\pagebreak[3]\hangindent=1cm\noindent11. S. di Serego Alighieri, G. Gavazzi, C. Giovanardi, et al., {\itshape{}Astron. Astrophys.} {\bf474}, 851 (2007).

\pagebreak[3]\hangindent=1cm\noindent12. G. Gavazzi, R. Giovanelli, M. P. Haynes, et al., {\itshape{}Astron. Astrophys.} {\bf482}, 43 (2008).

\pagebreak[3]\hangindent=1cm\noindent13. K. N. Abazajian, J. K. Adelman-McCarthy, M. A. Agueros, et al., {\itshape{}Astrophys. J. Suppl. Ser.} {\bf182}, 543 (2009).

\pagebreak[3]\hangindent=1cm\noindent14. I. D. Karachentsev and D. I. Makarov, {\itshape{}Astron. J.} {\bf111}, 794 (1996).

\pagebreak[3]\hangindent=1cm\noindent15. J. L. Tonry, A. Dressler, J. P. Blakeslee, et al., {\itshape{}Astrophys. J.} {\bf546}, 681 (2001).

\pagebreak[3]\hangindent=1cm\noindent16. H. Jerjen, B. Binggeli, and F. D. Barazza, {\itshape{}Astron. J.} {\bf127}, 771 (2004).

\pagebreak[3]\hangindent=1cm\noindent17. S. Mei, J. P. Blakesley, P. Cote, et al., {\itshape{}Astrophys. J.} {\bf655}, 144 (2007).

\pagebreak[3]\hangindent=1cm\noindent18. R. B. Tully, E. J. Shaya, I. D. Karachentsev, et al., {\itshape{}Astrophys. J.} {\bf676}, 184 (2008).

\pagebreak[3]\hangindent=1cm\noindent19. I. D. Karachentsev and O. G. Nasonova (Kashibadze), {\itshape{}Mon. Notic. Roy. Astron. Soc.} {\bf405}, 1075 (2010).

\pagebreak[3]\hangindent=1cm\noindent20. E. A. Evstigneeva, M. D. Gregg, M. J. Drinkwater, and M. Hilker, {\itshape{}Astron. J.} {\bf133}, 1722 (2007).

\pagebreak[3]\hangindent=1cm\noindent21. E. A. Evstigneeva, M. J. Drinkwater, C. Y. Peng, et al., {\itshape{}Astron. J.} {\bf136}, 461 (2008).

}

\end{document}